\documentstyle[twoside,fleqn,npb,epsfig,axodraw]{article}
\newcommand{\AmS}{{\protect\the\textfont2
  A\kern-.1667em\lower.5ex\hbox{M}\kern-.125emS}}

\begin{document}

\title{ $\null$\\
 \vskip - 1.6 cm  
{\small MSUHEP--21111 \hfill UCLA/02/TEP/34 \hfill SLAC--PUB--9573} \\
 $\null$ \\
The Di-Photon Background to a Light Higgs Boson at the LHC}

\author{Zvi Bern,\address{Department of Physics and Astronomy \\
                   UCLA, Los Angeles, CA 90095-1547, USA}%
    \thanks{Presenter
          at RADCOR 2002/Loops and Legs in Quantum Field Theory
                (September 2002, Kloster Banz, Germany).
          Research supported by the US Department of Energy under grant
          DE-FG03-91ER40662.}
   Lance Dixon\address{Stanford Linear Accelerator Center\\
                          Stanford University, Stanford, CA 94309, USA}%
     \thanks{Research supported by the US Department of Energy under contract
                 DE-AC03-76SF00515.}
   and
Carl Schmidt\address{Department of Physics and Astronomy \\
Michigan State University, East Lansing, MI 48824, USA}%
\thanks{Research supported by the US National Science Foundation under
grant PHY-0070443.}
}

\def\to{\rightarrow}
\def\lr{\leftrightarrow}
\def\Re{{\rm Re}}
\def\lsim{\buildrel < \over {_\sim}}
\def\gsim{\buildrel > \over {_\sim}}     
\def\qb{{\bar{q}}}
\def\alphas{\alpha_s}
\def\eps{\epsilon}
\def\e{\epsilon}
\def\mgg{M_{\gamma\gamma}}
\def\pt{p_{\rm T}}
\def\et{E_{\rm T}}
\def\etmax{E_{\rm T\,max}}
\def\etjet{E_{\rm T\,jet}}
\def\Rjet{R_{\rm jet}}
\def\ptmiss{/{\hbox{\kern-6pt $\pt$}}}
\def\Ord{{\cal O}}
\def\MS{\rm\overline{MS}}
\def\Nf{{N_{\! f}}}
\def\cm{{\cal M}}
\def\Sublead{{\rm \scriptscriptstyle SL}}
\def\Lead{{\rm \scriptscriptstyle L}}
\def\spa#1.#2{\left\langle#1\,#2\right\rangle}
\def\spb#1.#2{\left[#1\,#2\right]}
\def\lsim{\buildrel < \over {_\sim}}
\def\gsim{\buildrel > \over {_\sim}}     
\def\fig#1{fig.~{\ref{#1}}}
\def\Fig#1{Figure~{\ref{#1}}}
\def\eqn#1{eq.~(\ref{#1})}
\def\eqns#1#2{eqs.~(\ref{#1}) and~(\ref{#2})}
\def\draftnote#1{{\it #1}}

\begin{abstract}
Recent years have seen a significant advance in our ability to
calculate two-loop matrix elements.  In this talk we describe an
application of this breakthrough to improve our understanding of the
background to the search for a light Higgs boson at the LHC.  In
particular, we focus on the QCD corrections to the gluon fusion
subprocess $gg \to \gamma\gamma$, which forms an important component
of the background in the di-photon channel.  We find that
the $K$ factor for this subprocess is significantly smaller
than estimated previously.

\end{abstract}

\maketitle

\section{INTRODUCTION}

Many of the talks at this conference described the recent wonderful
advances in two-loop perturbative calculations (see
e.g. refs.~\cite{OtherTalks,AbilioTalk,AnastasiouTalk} and references
therein).  Here we outline the application of this breakthrough to a
phenomenological study~\cite{Hbkgd} of the background to the search
for a Higgs boson at the LHC in the di-photon mode.  In a
complementary talk at this conference, Abilio De
Freitas~\cite{AbilioTalk} described the calculation~\cite{GGGamGam} of
the two-loop matrix elements for gluon fusion into a photon pair
used in this study.

Arguably, the most pressing problem in particle physics today is the
origin of electroweak symmetry breaking.  Collider experiments over
the next decade should shed considerable light on this by searching
for the Higgs boson and measuring its physical properties.  No matter
what new physics lies beyond the Standard Model (SM), measurements of the
Higgs sector parameters will likely provide crucial clues to its
structure.

There are a few reasons to suspect that the mass of at least one Higgs
particle is quite light.  Most strikingly, the SM Higgs boson mass is 
bounded from above by precision electroweak measurements, 
$m_H \lsim 200$ GeV at 95\% CL.  There are also hints of a signal in 
the direct search in $e^+e^- \to HZ$ at LEP2, near the lower mass limit 
of 114~GeV. In the minimal supersymmetric theory the lightest Higgs mass
is $\lsim 135$ GeV.

For $m_H \lsim 140$ GeV, the preferred search mode at the LHC involves 
Higgs production via gluon fusion, followed by the rare decay into a pair of
photons~\cite{Higgsgammagamma}. (For a discussion of other useful
modes see {\it e.g.}  ref.~\cite{RSZ}.)  Although the branching ratio
is tiny, the di-photon mode is relatively clean due to the excellent
mass resolution of the LHC detectors, which will allow the
background to be measured experimentally and subtracted from a
putative signal~\cite{ATLAS,CMS}.  Nevertheless, it is still
important to have robust theoretical predictions in order to
systematically study the dependence of the signal relative to the
background to optimize Higgs search strategies. Since it
will take about two years of running at the LHC to extract the
$\gamma\gamma$ signal from the background, there is good motivation for
improving search strategies.

Given the intense theoretical effort that has gone into
two-loop calculations with more than a single kinematic variable, 
it is obviously beneficial to have concrete examples where the
new advances have already impacted phenomenology. The background to
Higgs decay discussed here is one such example.  Another recent
example was presented in the talk by Anastasiou~\cite{AnastasiouTalk},
where an exact calculation~\cite{AnastasiouMelnikov} of
next-to-next-to-leading-order (NNLO) inclusive Higgs
production~\cite{HKNNLO} was described.  The choice of Higgs physics
as among the first applications stems from both its importance for the
future of particle physics as well as the relative simplicity of the
infrared divergences encountered in the calculations. Once algorithms
are set up for dealing more generally with NNLO infrared divergent
phase space many more applications will certainly
appear~\cite{OtherTalks}.

\section{THE DI-PHOTON BACKGROUND}

The background to the Higgs search in the di-photon mode consists of
two pieces.  The `reducible' background arises when photons are faked 
by jets, or more generally by hadrons, especially $\pi^0$s. This 
background can be efficiently
suppressed by photon isolation cuts, where events are rejected based
on the hadronic energy near the
photons~\cite{ATLAS,CMS,Tisserand,Wielers,PiBkgd}.  The `irreducible'
background, which we focus on here, arises from the underlying QCD process
where quarks emit photons either directly or through fragmentation.

The process $pp \to \gamma\gamma X$ proceeds at lowest order via
quark annihilation, $q \bar q \to \gamma\gamma$.  The NLO
corrections to this subprocess have been incorporated into a
number of Monte Carlo programs~\cite{TwoPhotonBkgd1,TwoPhotonBkgd2};
the most up-to-date, {\tt DIPHOX}~\cite{DIPHOX}, also includes
fragmentation contributions.

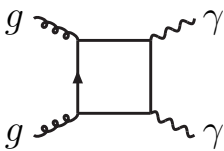
\begin{figure}[htb]
\begin{center}
\begin{picture}(70,40)(0,0)
\SetScale{.55}  
\SetWidth{2.0}
\Text(0,48)[l]{\Large$g$} \Text(0,5)[l]{\Large$g$}
\Gluon(20,90)(50,75){2.5}{3} \Gluon(50,25)(20,10){2.5}{3}
\ArrowLine(50,25)(50,75) 
\Line(50,75)(100,75) \Line(50,25)(100,25) \Line(100,75)(100,25)
\Photon(100,75)(130,90){2.5}{3} \Photon(100,25)(130,10){2.5}{3}
\Text(75,48)[l]{\Large $\gamma$}  \Text(75,5)[l]{\Large $\gamma$}
\end{picture}
\end{center}
\vskip -1 cm 
\caption{A leading order diagram contributing to gluon fusion 
into two photons.}
\label{LOGluonFigure}
\end{figure}

The largest of the contributions that have not yet been incorporated
into {\tt DIPHOX} are the NLO corrections to gluon
fusion into a di-photon pair.  Although the one-loop gluon fusion
contribution (\fig{LOGluonFigure}) is formally of higher
order in the QCD coupling than the tree-level process
$q \bar q \to \gamma\gamma$, 
it is enhanced by the large gluon distribution in the proton at small $x$,
so that it becomes numerically 
comparable~\cite{TwoPhotonBkgd1,TwoPhotonBkgd2,DIPHOX,HBkgdgammagamma,ADW}.  
To reduce the uncertainty on the total $\gamma\gamma$ production rate, a
calculation of the $gg \to \gamma\gamma$ subprocess at its
next-to-leading-order is required, even though it is formally N$^3$LO
as far as the whole process $pp \to \gamma\gamma X$ is concerned.
A number of other $\alpha_s^2$ and $\alpha_s^3$
contributions should eventually also be included,
although they are expected to be less significant~\cite{Hbkgd}.

\section{GLUON FUSION AT NLO}

\subsection{Matrix Elements and Singularities}

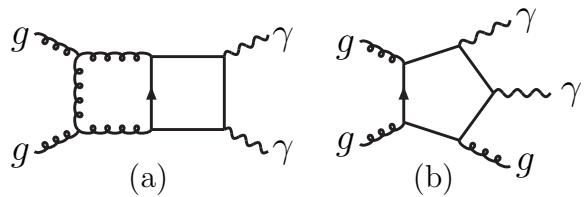
\begin{figure}[htb]
\begin{center}
\begin{picture}(220,50)(0,0)
\SetScale{.55}  
\SetWidth{2.0}
\Text(2,48)[l]{\Large $g$} \Text(2,4)[l]{\Large $g$}
\Gluon(20,90)(50,75){2.5}{3} \Gluon(50,25)(20,10){2.5}{3}
\Gluon(50,25)(50,75){2.5}{4} \Gluon(50,75)(100,75){2.5}{4}
\Gluon(100,25)(50,25){2.5}{4}
\ArrowLine(100,25)(100,75) 
\Line(100,75)(150,75) \Line(100,25)(150,25) \Line(150,75)(150,25)
\Photon(150,75)(180,90){2.5}{3} \Photon(150,25)(180,10){2.5}{3}
\Text(101,48)[l]{\Large $\gamma$}  \Text(101,5)[l]{\Large $\gamma$}
\Text(54,-5)[c]{\large(a)}
\SetOffset(-15,0)
\Text(140,45)[l]{\Large $g$} \Text(140,8)[l]{\Large $g$}
\Gluon(270,85)(300,70){2.5}{3} \Gluon(300,30)(270,15){2.5}{3}
\Gluon(372.88,0)(338.04,17.64){2.5}{3} 
\ArrowLine(300,30)(300,70) \Line(300,70)(338.04,82.36)
\Line(338.04,82.36)(361.55,50) \Line(361.55,50)(338.04,17.64)
\Line(338.04,17.64)(300,30)
\Photon(338.04,82.36)(372.88,100){2.5}{3} 
\Photon(361.55,50)(400.60,50){2.5}{3} 
\Text(208,55)[l]{\Large $\gamma$} \Text(225,27)[l]{\Large $\gamma$}  
\Text(208,0)[l]{\Large $g$}
\Text(177,-5)[c]{\large(b)}
\end{picture}
\end{center}
\vskip -.7 cm 
\caption{Sample NLO diagrams contributing to gluon fusion 
into two photons: (a) virtual and (b) real emission contributions. }
\label{NLOGluonFigure}
\end{figure}

The NLO correction to gluon fusion involves diagrams of the type
shown in \fig{NLOGluonFigure}.  The two-loop virtual contributions (a)
were recently computed~\cite{GGGamGam}, as summarized in 
the talk by De Freitas~\cite{AbilioTalk}. 
The real emission contributions (b) are obtained from a permutation
sum~\cite{TwoPhotonBkgd2} over contributions to the one-loop five-gluon
amplitude~\cite{GGGamGamG}.

Both the virtual and real corrections have been evaluated for zero
quark mass.   In the range of di-photon invariant masses 
$M_{\gamma\gamma}$ relevant for the light Higgs search (90--150 GeV)
this is an excellent approximation:  the masses 
of the five light quarks are negligible, while the top 
quark contribution is tiny for $M_{\gamma\gamma} \ll 2m_t \approx 350$ GeV.

In order to obtain a prediction for a physical cross section,
virtual and real corrections must be combined to cancel the infrared
divergences appearing in each term.  Because
$gg\to\gamma\gamma$ vanishes at tree-level, the divergences 
encountered in the NLO corrections to $gg \to \gamma\gamma X$
have the structure of a typical NLO QCD calculation, even
though two-loop matrix elements are involved.  We could therefore
employ the dipole formalism~\cite{CataniSeymour}.

\subsection{Photon Isolation}

To reduce fragmentation contributions and also to reject the
reducible background, photon isolation criteria are imposed.
There are two standard ways to do this:
(a) {\it standard} cone isolation where the amount of transverse hadronic
energy $\et$ in a cone of radius 
$R = \sqrt{(\Delta\eta)^2 + (\Delta\phi)^2}$ must be less than $\etmax$
and (b)  {\it smooth} cone isolation~\cite{Frixione} 
where the amount of transverse hadronic energy $\et$ in {\it all} cones 
of radius $r$ with $r < R$ must be less than a given function, for example:
\begin{equation}
\etmax(r) \equiv 
\pt(\gamma) \, \eps \, \biggl( { 1 - \cos r \over 1 - \cos R } \biggl) \,.
\end{equation}

The smooth cone is preferred theoretically because it
eliminates the fragmentation contributions.
However, it may be problematic experimentally, due to the finite width
of a photon's electromagnetic shower and the relatively
large granularity of the LHC calorimeters~\cite{Wielers}.  We have
implemented both isolation algorithms.  For the standard
cone we use {\tt DIPHOX}~\cite{DIPHOX} to obtain all
contributions except the gluon fusion ones; for the smooth cone
we constructed an independent NLO program for all pieces.

\section{RESULTS FOR DI-PHOTON BACKGROUND}

\begin{figure}
\includegraphics[width=7.cm]{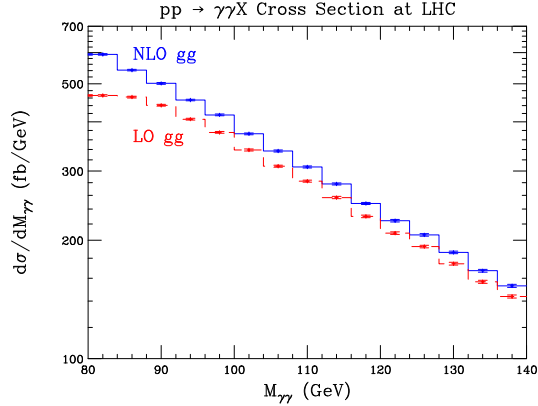}
\vskip -.8 cm 
\caption{\label{fig2_rp4} Total $pp \to \gamma\gamma X$ production at
NLO, including NLO $q\qb\to\gamma\gamma$ and fragmentation
contributions, with the gluon fusion subprocess treated at LO (lower)
and at NLO (upper). In these plots the
renormalization and factorization scales are $\mu_R = \mu_F =
\mgg/2$, using MRST99 set 2 partons. A standard photon isolation criterion
is used with $R=0.4$, $\etmax = 15$~GeV. }
\end{figure}

\Fig{fig2_rp4} illustrates the shift in the total NLO $pp \to
\gamma\gamma X$ production rate, for the case of standard cone
isolation, due to the gluon fusion subprocess.  The lower curve in the
plot is obtained from {\tt DIPHOX}~\cite{DIPHOX}, incorporating the
gluon fusion subprocess only at its leading order.  The upper curve
includes the NLO contributions to gluon fusion.  The
increase in the total irreducible $\gamma\gamma$ background which
results from replacing the LO gluon fusion quark box by the NLO
computation is a relatively modest one, except at the lowest invariant
masses which are not relevant for SM Higgs searches.  For
the most interesting mass range 115 GeV $< m_H <$ 140 GeV, the overall
effect on the square root of the background is under 5\%.  Thus, this
subprocess can be considered to be under adequate theoretical control.
A more detailed discussion of these results may be found in 
ref.~\cite{Hbkgd}.

Prior to the NLO calculation, some experimental studies had
used the $K$ factor (ratio of NLO over LO cross section) for Higgs
production by gluon fusion to estimate the $K$ factor for
$gg\to\gamma\gamma$.  After all, both $gg\to H$ and
$gg\to\gamma\gamma$ produce a colorless system from a
$gg$ initial state.  Actually, the background $K$ factor is 
significantly smaller, only about $65\%$ of the signal $K$ factor
for a broad range of Higgs masses.  
This difference is partially due to a short-distance
renormalization contribution that only affects the Higgs
production process.  The remainder can be traced to
the heaviness of the top quark in the Higgs production loop diagram:
the background process involves light quarks and is "softer" when an extra
gluon attaches to the loop~\cite{Hbkgd}.

\section{STATISTICAL SIGNIFICANCE OF HIGGS SIGNAL AND OUTLOOK}

\vskip -.1 cm 
To study the statistical significance of the signal, we
implemented the gluon fusion production of the SM Higgs
boson at NLO~\cite{NLOHiggs,NLOHKfactor}, followed by decay to
$\gamma\gamma$, with a branching ratio obtained from the program 
{\tt HDECAY}~\cite{HDECAY}.
As is standard for a light Higgs boson, we work in the heavy top quark limit,
for which an effective $Hgg$ vertex~\cite{HggVertex} suffices to
describe the production process at low Higgs transverse momenta.

The interference between the Higgs signal and background 
is rather tiny, mainly because the Higgs resonance in this
mass range is extremely sharp, but also because of
properties of the background amplitudes~\cite{Hbkgd}.
Thus one can consider the signal and background cross sections separately.

\subsection{Effects of Varying Photon Isolation}

Armed with a reliable theoretical calculation, one can systematically
study the effects on both signal and background of varying the photon
isolation criteria. One can also search for the best kinematic cuts
and variable choices for enhancing the signal over the background.  

\begin{figure}
\includegraphics[width=7.cm]{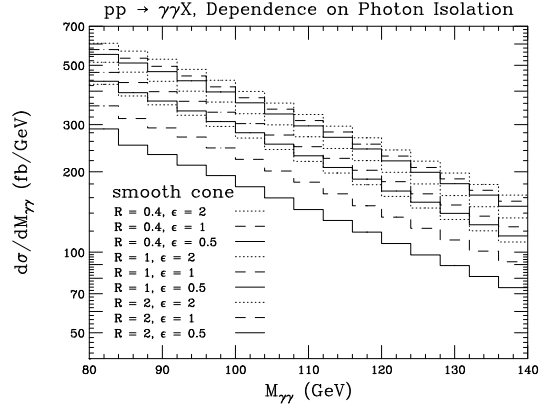}
\vskip -.8 cm 
\caption{\label{fig6} Dependence of $pp\to\gamma\gamma X$ at the LHC 
on photon isolation cuts, for a set of smooth cone isolation
parameters, $R$ and $\eps$.  All plots are for MRST99 set 2 partons, and 
$\mu_R = \mu_F = 0.5 \mgg$. }
\end{figure}

For example, \Fig{fig6} shows the dependence
of the $pp\to\gamma\gamma X$ production rate at the LHC on the
parameters of the smooth isolation cone criterion~\cite{Hbkgd}.  
As the isolation becomes more severe, {\it i.e.}
$R$ is increased or $\epsilon$ is decreased, the direct
$pp\to\gamma\gamma X$ background becomes more suppressed.  The large
sensitivity to these parameters is due to the $q\gamma$ collinear
singularity in the NLO $q\bar q \to \gamma\gamma X$ cross section.
Since the QCD radiation in Higgs production has no such singularity,
it should be uncorrelated with the photon directions, and
so the signal is less sensitive to the isolation criterion.

A better way to suppress the QCD background is with a jet veto.
At the NLO parton level, at least for direct processes, a jet veto 
corresponds closely to increasing the cone size.  However,
transverse energy can be forbidden into a smaller area (the jet cone
size), for the same amount of suppression at NLO.  Hence the jet veto should
be better behaved theoretically.  It should also have significantly
better experimental properties since less signal is lost due to
detector noise, overlapping events, {\it etc.}

In either of these cases, even though the background falls
significantly, it turns out that $S/\sqrt{B}$ is rather insensitive to
the tighter cuts~\cite{Hbkgd}.  One can improve the situation
slightly by taking into account information about the rapidity difference
between the two photons~\cite{Hbkgd}.
It also may be possible to find better variables characterizing the
hadronic energy flow in the events, to reduce the number of
signal events cut out.

Studies making use of hadronic energy flow would need to be carried out
with a more realistic simulation than the parton-level one outlined here.  
In particular, effects of instrumental noise and overlapping events
should be included~\cite{Wielers}.  It would also be very important to
include a detailed study of the reducible $\pi^0$ background
contributions~\cite{Tisserand,Wielers,PiBkgd}.
Finally, other NNLO contributions could be added as they become available.

The improved calculation of the di-photon background to Higgs decay at
the LHC described in this talk is the first collider physics
application of the new two-loop QCD amplitudes depending on more than
a single kinematic variable.  In the near future, many
more phenomenological applications should be forthcoming.


\end{document}